\documentclass[reprint,amsmath,amssymb,aps,superscriptaddress,nofootinbib,twocolumn,10pt]{revtex4-1}
\usepackage[colorlinks,linkcolor=blue,citecolor=blue,urlcolor=blue]{hyperref}
\usepackage{graphicx,epstopdf}
\usepackage{dcolumn}
\usepackage{subfigure}
\usepackage{bm}
\usepackage{amsmath,amsfonts,amssymb,graphicx,multirow,dcolumn,bm,latexsym,soul,nicefrac}
\usepackage{acronym}
\usepackage{enumitem}
\usepackage{array}
\usepackage{multirow}
\usepackage{appendix}
\usepackage{longtable}
\usepackage{footnote}
\usepackage{amssymb}

\newcommand{\SPA}{School of Physics and Astronomy, Sun Yat-sen University (Zhuhai Campus), Zhuhai 519082, China.}

\begin{document}

\title{Statistic threshold of distinguishing the environmental effects and modified theory of gravity with multiple massive black-hole binaries}

\author{Xulong Yuan}
\email{yuanxlong@mail2.sysu.edu.cn}
\affiliation{\SPA}

\newacro{GR}{general relativity}
\newacro{GW}{gravitational wave}
\newacro{MBH}{massive black hole}
\newacro{MBHB}{massive black hole binary}
\newacro{EMRI}{extreme mass-ratio inspiral}
\newacro{IMRI}{intermediate mass-ratio inspiral}
\newacro{ppE}{parameterized post-Einsteinian}
\newacro{BH}{black hole}
\newacro{NS}{neutron star}
\newacro{BS}{boson star}
\newacro{ECO}{exotic compact object}
\newacro{LVK}{LIGO-Virgo-KAGRA}
\newacro{BBH}{binary black hole}
\newacro{BNS}{binary neutron star}
\newacro{NSBH}{neutron star-black hole}
\newacro{SIQM}{spin induced quadrupole moment}
\newacro{ASD}{amplitude spectral density}
\newacro{PSD}{power spectral density}
\newacro{SNR}{signal-to-noise ratio}
\newacro{PE}{parameter estimation}
\newacro{PN}{post-Newtonian}
\newacro{FIM}{fisher information matrix}
\newacro{TDI}{time delay interferometry}
\newacro{ISCO}{innermost stable circular orbit}
\newacro{PDF}{probability distribution function}
\newacro{EPS}{extended Press and Schechter}
\newacro{LISA}{Laser Interferometer Space Antenna}
\newacro{ppE}{parameterized post-Einstein}
\newacro{DF}{dynamical friction}
\newacro{DM}{dark matter}
\newacro{IMRI}{intermediate mass-ratio inspiral}
\newacro{Df}{dynamical friction}
\newacro{DM}{dark matter}
\newacro{MBH}{massive black hole}
\newacro{RS}{Randall and Sundrum}
\newacro{EdGB}{Einstein-dilaton-Gauss-Bonnet}
\newacro{EA}{Einstein-Æther }
\newacro{ROC}{Receiver Operating Characteristic Curve}
\newacro{AUC}{area under curve}

\def\mrd{\mathrm{d} }
\def\pd{\partial}
\def\m{\tx{m}}
\def\kg{\tx{kg}}
\def\s{\tx{s}}
\def\rsp{r_\tx{sp}}
\def\d{\Delta }
\def\dt{\delta}
\def\hz{\text{Hz}}
\def\be{\begin{equation}}
	\def\ee{\end{equation}}
\def\ba{\begin{eqnarray}}
	\def\ea{\end{eqnarray}}
\def\nn{\nonumber}
\newcommand{\tx}[1]{\text{#1}}
\newcommand{\scf}[1]{\times 10^{#1}}
\newcommand{\td}[1]{\tilde{#1}}
\newcommand{\mc}[1]{\mathcal{#1}}
\newcommand{\secref}[1]{Sec.~\ref{#1}}
\newcommand{\eqnref}[1]{(\ref{#1})}
\newcommand{\figref}[1]{Fig.~\ref{#1}}
\newcommand{\appref}[1]{Appendix~\ref{#1}}
\renewcommand{\eqref}[1]{Eq. (\ref{#1})}

\newcommand{\tabref}[1]{\textbf{Table.~\ref{#1}}}

\begin{abstract}
In future space-borne \ac{GW} observations, matter arround the sources might influence the evolution and \ac{GW} signals from \ac{BBH} inspirals, which can be mistaken as deviations from \ac{GR}. Former research \cite{yuan2024} proposed a statistic $F$ that characterizes the dispersion of measured parameters to distinguish environmental effect(\ac{DF} from \ac{DM} spike) and theory of modified gravity effect(varying $G$). In this work we use the statistic to distinguish other couples of effects with \ac{GW} corrections at $-4$ PN order: \ac{DF} from \ac{DM} spike and the extra dimension theory, additionally try to determine the distinguishing threshold in more reasonable way to avoid arbitrariness, especially when the two effects to compare have more overlap in the $F$ distribution. Sources of different astronomical models are also considered, and two effects are still distinguishable but not much as in former work\cite{yuan2024}, so the threshold should be carefully selected. Following these procedures, we finally obtain the statistic thresholds of distinguishment between the three effects with \ac{GW} corrections at $-4$ PN order: \ac{DF} from \ac{DM} spike, the extra dimension theory, and varying $G$ theory. The method can be used to distinguish other effects among environmental effects and modified theories of gravity effects with the detections of \ac{GW} events.
\end{abstract}

\maketitle

\section{Introduction}
\label{intro}
Several space-borne gravitational wave detectors have been designed and in schedule to operate in the coming 2030s including TianQin \cite{Luo:2015ght,TianQin:2020hid,Luo:2025sos}, LISA \cite{Danzmann:1997hm, LISA:2017pwj}, and Taiji\cite{Hu:2017mde}. All of them focus on the milihertz frequency band of \ac{GW}. One of their most important \ac{GW} sources are inspirals of \acp{MBHB} \cite{Klein:2015hvg,Wang:2019ryf,Feng:2019wgq}. Massive black holes are expected to harbour at the center of galaxies, and they may form coalescence after galaxies merger. The longer observation duration of \acp{MBHB} and larger \ac{SNR} with space-borne detectors will reflect the evolution of \acp{MBHB} on the path to merger, which helps enhance the knowledge of fundamental physics\cite{Gair:2012nm, Barausse:2020rsu, LISA:2022kgy,Perkins:2020tra}, astrophysics \cite{LISA:2022yao, Baker:2019pnp} and cosmology \cite{Tamanini:2016zlh, Zhu:2021aat,LISACosmologyWorkingGroup:2022jok, Caldwell:2019vru,Zhu:2021aat,Zhu:2021bpp}, especially the nature of gravity in the strong field regime\cite{Shi:2019hqa,Shi:2022qno,Kong:2024ssa,Tan:2024utr,Rahman:2022fay}.


Most analysis have considered vacuum \ac{GW} sources. However, many effects may cause deviation in \ac{GW} signals that are similar to violation of \ac{GR}, which should be recognized and distinguished from the violation. They can be categorized as noise systematics, waveform systematics, and astrophysical aspects\cite{Gupta:2024gun,Dhani:2024jja,Lau:2024idc,Garg:2024qxq,Chandramouli:2024vhw}, and in this article we focus on the latter two factors.

On one hand, modeling of \acp{GW} from compact binaries can introduce waveform systematics. Some of them are within the calculation of \ac{GR} waveform, while others are beyond \ac{GR}. Various modified theories of gravity have been proposed since the birth of Brans-Dicke theory as the first one. They may bring waveform corrections and a number of them have been studied in different approaches. Based on the \ac{PN} method as well as an extension \ac{ppE} formalism\cite{bppe,ppewv}, the waveform correction of theories of modified gravity have been derived, such as the Scalar tensor\cite{PhysRevD.65.042002,esbgtl}, \ac{EdGB}\cite{PhysRevD.85.064022}, DCS\cite{PhysRevD.94.084002,PhysRevLett.109.251105}, \ac{EA}\cite{Hansen:2014ewa}, Khronometric\cite{Hansen:2014ewa}, Noncommutative\cite{Kobakhidze:2016cqh}, varying $G$\cite{gdxw} and extra dimension\cite{PhysRevD.83.084036,PhysRevD.94.084002}. Parameter estimation of those theories have been analyzed considering ground-based and space-based detectors\cite{PhysRevD.94.084002,Zhang:2017sym,Hansen:2014ewa,gdxw,Shao:2020shv,PhysRevD.85.024041,PhysRevD.96.084039}.

On the other hand, astrophysical effects are also potential to influence the \acp{GW} from the compact binaries, usually they are catogorized as effects during the propagation and effects around source. Matter laying on the path can leads to gravitational lensing\cite{Lin:2023ccz,Tambalo:2022wlm,Caliskan:2022hbu}. Surrounding the source, there could be dark matter halo, the accretion disk, and the third body\cite{Camilloni:2023xvf,Tahelyani:2024cvk,AbhishekChowdhuri:2023rfv}. Binary immersed in the environments may experience accretion, gravitational pull and dynamical friction\cite{Barausse:2014tra}. Environmental effects near the source will generate modifications to waveform that may mimic modified theory of gravity effects. 

One of the environmental effects is the dark matter spike \cite{dmagc}, which is predicted to exist around the massive black holes in the center of the galaxies.
Though it has not been confirmed by current observations, future \ac{GW} detectors are expected to verify the existence of it or not. Considering dark matter mini-spikes, gravitational pull and dynamical friction force will both introduce waveform corrections\cite{eda,pmm}. Eccentricity is also taken into account later\cite{gyy}.
For \acp{MBHB} immersed in \ac{DM} spike\cite{Barausse:2014tra,constr}, dynamical friction will dominate the evolution and thus it should be the leading order effect in waveform correction.
From the observed events of LIGO, the density is constrained lately\cite{fclg}.

Former research has considered the bias of parameter posterior in Bayersian inference of \ac{IMRI} or \ac{EMRI} systems introduced by both modified theories of gravity and environmental effect\cite{Speri:2022upm,mdme}, and Bayes factors considering some environmental effects are also calculated\cite{Cole:2022yzw}. Speri $et$ $al$ \cite{Speri:2022upm} found the posterior of primary mass and spin of injected migration signal, analyzed with vacuum template and GR deviation template are both biased from the true distribution, however, the posterior with GR deviation template is minor. Thus, environmental effects like migration could lead to false detection of a GR deviation.

Consequently, some method to distinguish environmental effects and similar modified gravity effects are needed. Recently, Yuan $et$ $al$ \cite{yuan2024} considered groups of sources for different \ac{MBH} models, deriving usual $\dot{G}$ parameters of the sources from varying $G$ theory and from variable $\dot{G}$ distribution corresponding to \ac{DF} from \ac{DM} spike with power index $\gamma={3\over 2}$, where the \ac{GW} phase corrections of both \ac{DM} and varying $G$ effects are introduced at $-4$ PN order. Adopting Fisher analysis, they obtained the detection precisions of $\dot{G}$ from the sources' \ac{GW} signals with the space-borne \ac{GW} detector Tianqin, then used the $\dot{G}$ and $\delta\dot{G}$ data to construct a statistic $F$ for each group that reflects the dispersion of mean value and the uncertainty around the mean. They found the statistics $F$ for $\dot{G}$ distribution corresponding to \ac{DF} of \ac{DM} spike are much greater than those of the usual constant $\dot{G}$ case predicted by varying $G$ theory because \ac{DM} effect, depending on local \ac{DM} density is much more diverse. The $F$ statistic intervals of two effects hardly have overlap, devided by value of about $F=10^{10}$. Therefore, environmental effect and the selected modified gravity effect are distinguishable.

Although sometimes the distinguishing boundary can be found directly with the eyes, like the boundary of statistic $F$ in \cite{yuan2024}, it is not the case in general. In this article, we will apply the statistic $F$ to distinguish environmental effect from other modified gravity effect, i. e. the extra dimension theory and try to determine the proper threshold of statistic $F$ via \ac{ROC} curve method, considering sensitivity and specificity of hypothesis testing when the two effects have wider overlaping interval of $F$ distributions. Since these two effects are more similar than former result\cite{yuan2024}, it is more important to determine the distinguishing threshold. We will also implement the $F$ statistic to distinguishment between two modified theories of gravity. Finally together with former result in \cite{yuan2024}, we will compare three envolved effects directly, thus future observed statistic $F$ can map to corresponding effect at once. These methods can further be adopted in the distinguishment between other effects with the same leading order \ac{GW} phase corrections, whether of environmental origin or theoretic origin.

This article is organized as follows. In Sect.~\ref{sec:2}, we revisit the \ac{ppE} formalism with higher modes correction, and based on it, introducing the waveform corrections of \ac{DF} from \ac{DM} spike, the extra dimension theory and the theory of varying $G$. These are the effects to be distinguished later. In Sect.~\ref{sec:3}, we utilize the statistic $F$ by Yuan $et$ $al$ \cite{yuan2024} to distinguish \ac{DF} from \ac{DM} spike and the extra dimension theory ettects, and exploit the \ac{ROC} curve method to determine a proper threshold of $F$ when there is more overlap between the two $F$ distributions. After conversing to varying $G$ effect, we repeat this process to distinguish between the three effects, combined with the results of Yuan $et$ $al$ \cite{yuan2024}. Finally we draw a conclusion in In Sect.~\ref{concl}. We adopt $G=c=1$ in this work.
\section{Waveform model}\label{sec:2}

A lot of modified theories of gravity have been propoesed for various purpose, and a majority of them introduce waveform corrections that can be characterized within the \ac{ppE} framework. Similarly, astrophysical environment can influence the binary's orbital evolution and the emitting \ac{GW}, within \ac{ppE} formalisn as well. In this section, we will introduce the \ac{ppE} framework considering higher modes. With this then we will revisit waveform corrections of the environmental effect of our interest: \ac{DF} of \ac{DM} spike, and waveform corrections of two modified theories of gravity: the varying $G$ theory and the extra dimension theory. All of their phase corrections are introduced at $-4$ PN order. In the following part of this section we will introduce these effect. Since this work extends the work of Yuan $et $ $al$\cite{yuan2024}, we adopt the same parameters for setup process with it.

%

\subsection{The \ac{ppE} framework}
\label{ppes}

The \ac{ppE} formalism was introduced to describe the leading order corrections in a parametric form\cite{bppe}, with which we consider environmental effect or modified theory of gravity in \ac{GW}. When the masses of two components are highly asymmetric, higher modes of their \acp{GW} become important. Taking into account higher modes, the \ac{ppE} formalism of \ac{GW} correction is\cite{gjmfk}
\begin{equation}
	\td{h}(f)=\sum_{lm}\td{h}_{lm}(f)=\sum\td{h}^\text{GR}_{lm}(f)e^{i\delta\Psi_{lm}}.
\end{equation}
Here we have ignored amplitude correction because it is nearly negligible and the phase correction is
\begin{equation}
	\delta\Psi_{lm}=\beta_{lm}u^{b_{lm}},
\end{equation}
where $u=(\pi\mc{M}_cf)^{1/3}$ is the relative velocity of the binary components, and $\mc{M}_c=(m_1m_2)^{3/5}/(m_1+m_2)^{1/5}$ is the chirp mass for the binary with component masses $m_1$ and $m_2$. The \ac{ppE} parameters of higher modes are related with those of 22 mode through
\begin{equation}
	\beta_{lm}=\left(\frac{2}{m}\right)^{\frac{b_{lm}}{3}-1}\beta_{22},~~~b_{lm}=b_{22}.
\end{equation}
In this article, we use IMRPhenomXHM\cite{phexhm} to generate the \ac{GR} waveform with higher modes. 

\subsection{\ac{DF} of \ac{DM} spike}

Due to the adiabatic growth of \ac{BH}, the surrounding \ac{DM} density increases and forms a spike in \ac{DM} halo\cite{dmagc}. The distribution of \ac{DM} spike is often steeper close to the center, which can be expressed with a power-law profile
\begin{equation}
	\rho(r)=\rho_0\left(\frac{r_0}{r}\right)^\gamma
\end{equation}
where $r$ is the distance to the center of mass of the binary. Note that $\rho_0$ is the density at $r_0$. The range of power index $\gamma$ is expected to be $[2.25,2.5]$ for steep spike\cite{dmagc}. We ignore the inner radial cutoff here because it will be the \ac{ISCO} of the binary. 

When the binary components are moving in the \ac{DM} medium, they will invoke over-density following them, or the so-called 'wake' which exerts friction force\cite{dfi} (dubbed \ac{DF} force) on each component, and \ac{DF} of the \ac{DM} spike introduces extra energy loss in addition to gravitational radiation in the energy balance equation. The corresponding phase correction is then\cite{yuan2024,pmm,constr}
\begin{equation}
	\delta\Psi_\text{DF}=-\frac{75I\pi\rho_0r_0^\gamma}
	{256(\gamma-8)(2\gamma-19)}Ku^{2\gamma-16},
\end{equation}
where the factor $K$ is
\begin{eqnarray}
	K&=&(m_1^2m_2^{-\gamma-1}+m_2^2m_1^{-\gamma-1})\mc{M}_c\eta^{-\frac{2\gamma+2}{5}}\nn\\
	&=&\frac{
		\left(1+\sqrt{1-4\eta}\right)^{3+\gamma}+\left(1-\sqrt{1-4\eta}\right)^{3+\gamma}}
	{\mc{M}_c^{\gamma-2}\eta^{\frac{4\gamma}{5}+2}2^{\gamma+3}}\label{equ:phase}
\end{eqnarray}
and the values of $r_0$ and $\rho_0$ is determined by the \ac{DM} mass(which is related to the critical density of our universe $\rho_c=9.73\times 10^{-27}\text{kg}/\text{m}^3$ \cite{WMAP1OB} and the \textcolor{black}{redshift} $z$)  and \ac{BH}s' radius of gravitational influence as\cite{yuan2024,pmm}:
\begin{eqnarray}
	\rho_0&=&200 \rho_{cm}5^\gamma\left(\frac{N}{2}\right)^\frac{\gamma}{3-\gamma}\label{equ:rho0}\\
	r_0&=&\left(\frac{M(3-\gamma)}{2\pi\rho_0}\right)^\frac{1}{3}5^\frac{\gamma-3}{3}.\label{equ:r0}
\end{eqnarray}
$N$ is generally assumed to be in the range of  $10^3-10^6$\cite{eda,mdme,galdy,Sesana:2014bea,yuan2024}. We take $N=10^6$ in this article unless specially noting. $I$ is \ac{DF} factor depending on the velocities of components with respect to the medium\cite{kim2}. We adopt $I\cong3$ as in \cite{pmm,yuan2024,constr}.

So the \ac{ppE} parameters of 22 mode \ac{DF} phase correction is
\begin{eqnarray}
	\beta_{22}&=&-\frac{75I\pi\rho_0r_0^\gamma}
	{256(\gamma-8)(2\gamma-19)}K,\\
	b_{22}&=&2\gamma-16.
\end{eqnarray}

Treating $\rho_0$ as a parameter to be measured, from \eqref{equ:rho0} and \eqref{equ:r0} the expression $\rho_0r_0^\gamma$ can be furthur symplified
\begin{equation}
	\rho_0r_0^\gamma=(2\pi)^{-\frac{\gamma}{3}}5^\frac{\gamma(\gamma-3)}{3}
	(3-\gamma)^\frac{\gamma}{3}\mc{M}_c^\frac{\gamma}{3}
	\eta^{-\frac{\gamma}{5}}\rho_0^{1-\frac{\gamma}{3}}.
\end{equation}
Taking $\gamma=\frac{3}{2}$, then $b=-13$ so
\begin{equation}
	\delta\Psi_\text{DF}=-\frac{15\sqrt[4]{5}}{851968}\sqrt{\frac{3\pi\rho_0}{2}}
	I\mc{M}_c\eta^{-\frac{7}{2}}H(\eta)u^{-13}\label{dfxz}
\end{equation}
is $-4$ PN correction and the factor $H(\eta)$ is
\begin{equation}
	H(\eta)=\left(\frac{1}{2}+\frac{\sqrt{1-4\eta}}{2}\right)^\frac{9}{2}
	+\left(\frac{1}{2}-\frac{\sqrt{1-4\eta}}{2}\right)^\frac{9}{2}
\end{equation}
Although $\gamma={3\over 2}$ is not within the general range of $[2.25,2.5]$, it is also considered in other work\cite{pmm,yuan2024}.

\subsection{Extra dimension effect}\label{eds}

One of the modified gravity theories that predicts the same leading order phase correction, i.e. $b=-13$ as \ac{DF} from \ac{DM} spike with power index $\gamma={3\over 2}$ is \ac{RS}-II braneworld model. According to \ac{RS}-II theory, mass of \ac{BH} will decrease with time, $\dot{m}\ne0$ due to gravitational leakage into the bulk\cite{Johannsen_2009}, thus introduce phase correction as
\ba 
\dt\Psi_\tx{ED}=\beta_{22,\mathrm{ED}}u^b\label{dped}
\ea
where\cite{PhysRevD.83.084036}
\ba 
\beta_{22,\mathrm{ED}}=\frac{25}{851968}\left(\frac{dm_1}{dt}+\frac{dm_2}{dt}\right)\frac{3 - 26\eta+34\eta^{2}}{\eta^{2/5}(1 - 2\eta)}\label{edxw}
\ea
and
\ba 
\frac{dm}{dt}=-2.8\times 10^{-7}\left(\frac{1M_{\odot}}{m}\right)^{2}\left(\frac{\ell}{10\mu\mathrm{m}}\right)^{2}\frac{M_{\odot}}{\mathrm{yr}}
\ea
is the evaporation rate of a \ac{BH} $m$ and $\ell$ is the size of the large extra dimension.

Current table-top experiments places the bound $\ell<14\mu\m$, given by test of the gravitational inverse-square law. Yunes $et$ $al$ \cite{PhysRevD.94.084002} showed aLIGO observation of events GW150914 and GW151226 requires $\ell \lesssim 10^{9}\mu\m$. Considering future ground-based detectors, the constraint can only be improved by three orders of magnitude\cite{PhysRevD.96.084039}. However, space-borne detectors may give stronger constraint of $\ell$, and Yagi $et$ $al$ \cite{PhysRevD.96.084039} predicted an \ac{EMRI} system $(10,10^5)M_\odot$ at SNR $=100$ observed by LISA for one year can improve it to $\ell\le 10^3\mu\m$.

\subsection{The theory of varying $G$}
Another modified theory of gravity with \ac{ppE} correction parameter $b=-13$ is the theory of varying G theory. As mentioned before, the distinguishment of \ac{DM} spike \ac{DF} effect and varying G theory has been given by Yuan $et$ $al$\cite{yuan2024}, and we extend their analysis to compare all three effect, including distinguishing two modified theories of gravity.

In \ac{GR}, $G$ is the gravity constant independent of spacetime.
While in the varying $G$ theory, it will vary with rate $\dot{G}$\cite{gdxw},
which causes a phase correction \cite{ppewv} as

\begin{equation}
	\delta\Psi_{\dot{G}}=-\frac{25\mathcal{S}}{851968}\dot{G}\mc{M}_c(\pi\mc{M}_cf)^{-{13\over 3}}.\label{sbyl}
\end{equation}
\textcolor{black}{so the corresponding \ac{ppE} phase parameters of 22 mode are}
\begin{eqnarray}
	\beta_{22}&=&-\frac{25\mathcal{S}}{851968}\dot{G}\mc{M}_c,\\
	b_{22}&=&-13.
\end{eqnarray}

For general binaries, the factor $\mathcal{S}$ in $\beta_{22}$is\cite{gdxw,pdppe}
\begin{equation}
	\mathcal{S}=11-\frac{35}{2}(s_1+s_2)+\frac{41}{2}\sqrt{1-4\eta}(s_1-s_2),
\end{equation}
where $s_{1,2}$ are the sensitivities for the components.
Specifically, for \acp{BH}, we will have $s_1=s_2=\frac{1}{2}$, and thus obtain the factor $\mathcal{S}=-\frac{13}{2}$
\cite{sbdrvg}.

Above all, both the leading order correction of the selected extra dimension effect and of varying-G theory is -4 PN,
the same order \textcolor{black}{as the \ac{DF} correction} with $\gamma=\frac{3}{2}$.
Thus these effects will have degeneracy in \ac{GW} phase correction, and if we find the -4 PN deviation in the future observations especially in a single event,
it will be difficult to account for it. It can either originate from environmental effect of \ac{DM}, or from the two modified theories of gravity.

\section{Distinguishing \ac{DM} and extra dimension effect}\label{sec:3}
In this section, we will apply the statistic $F$ for the distinguishment of \ac{DM} and extra dimension effects.

\subsection{Relation of the two waveform corrections}\label{rwc}
In the usual \ac{RS}-II braneworld theory, the size of the large dimension $\ell$ is a constant independent of lacation. If the actual wavefrom correction is caused by \ac{DF} from \ac{DM} spike arrond the source, the correction can alternatively be explained by the extra dimension theory with $\ell$ which, however, is no longer a constant.

Equating the waveform correction of \ac{DF} from \ac{DM} spike \eqref{dfxz} and that of extra dimension \eqref{dped}, directly yields the extra dimension parameter $\ell$ corresponding to \ac{DF} effect from \ac{DM} spike,
\ba 
\ell&=&10\sqrt{75I\rho_0r_0^\beta \pi \mc{M}_c^2K\over 256(2\beta-19)(\beta-8)\times 1.3\times 10^{-24}   H}.\label{dfl}\\
H&=&\frac{3 - 26 \eta + 34 \eta^2}{\eta^{\frac{2}{5}} (1 - 2 \eta)}\left(\frac{1}{m_1^2} + \frac{1}{m_2^2}\right)\nn
\ea
Because there exists a zero point in \eqref{dfl}:
\ba 
\eta_0={13-\sqrt{67}\over 34}\approx 0.14
\ea
of factor $3-26\eta+34\eta^2$ in \eqref{dfl}, when $\eta>\eta_0$, there is no  $\ell$ corresponding to \ac{DM} effect, and \ac{GW} with \ac{DM} effect from sources within this range will not be messed as extra dimension origin, meaning these two effects are totally distinguishable. However, when 
\ba 
\eta<\eta_0,\label{etas}
\ea 
waveform correction of \ac{DM} can be mimicked by extra dimension effect through \eqref{dfl}, and later we will use the $F$ statistic to distinguish them.

\begin{figure}
	\begin{center}
		\includegraphics[scale=0.55]{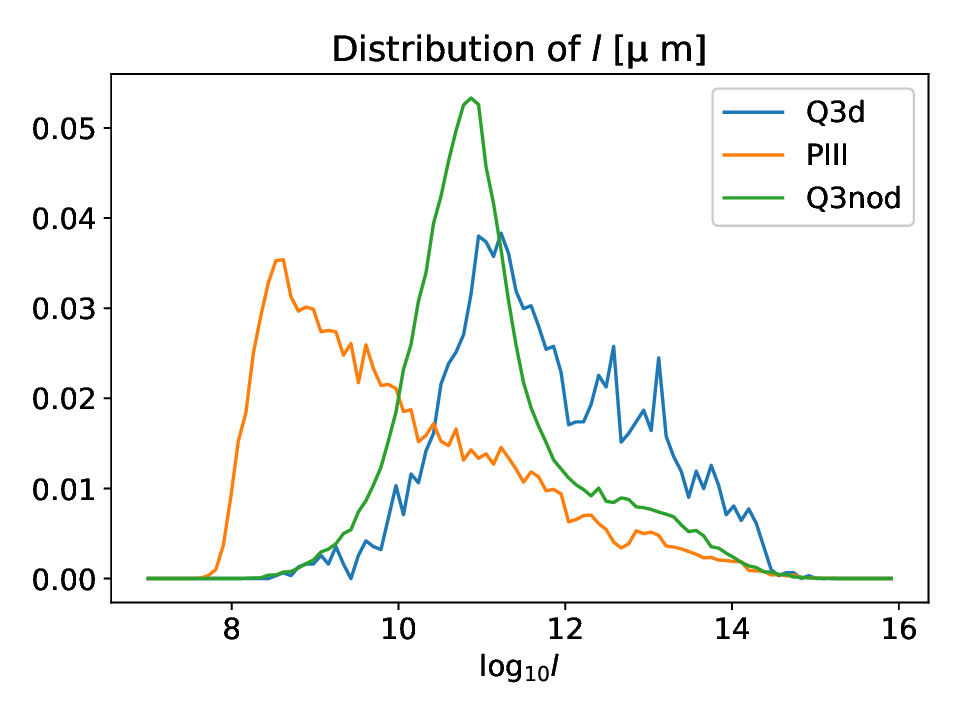}
		\caption{The $\ell$ distributions corresponding to \ac{DF} effect from \ac{DM} spike of the three astrophysical models}\label{trhgded}
	\end{center}
\end{figure}

Future \ac{GW} observation data will be used to see if there is violations of \ac{GR}, and sources at different locations enable us to break the degeneracy in waveform correction due to similar effects. In this article, we will consider three different models \cite{Barausse:2012fy,Sesana:2014bea,Antonini:2015sza,Klein:2015hvg}in our work,
two heavy-seed models ``Q3\underline{~}d'' and ``Q3\underline{~}nod'', and a light-seed model ``popIII''. Each model has 1000 mock catalogues of sources containing parameters $(z,m_1,m_2,\chi_1,\chi_2,\iota)$ for a five-year observation with TianQin\cite{Wang:2019ryf},
and their total events number are 18112, 271444, and 56618, respectively. We only consider events with \ac{SNR}$>8$, and satisfying \eqref{etas} in this case.

From \eqref{dfl}, $\ell$ values corresponding to the \ac{DF} effect from \ac{DM} spike of the three astronomical models Q3d, PIII and Q3nod is ploted in \figref{trhgded}.

\subsection{Fisher information matrix}

Another ingredient of the statistic $F$ is the detection precision at some value of a parameter. In this work, we use Fisher information matrix method to derive it. In large \ac{SNR} limit, the posterior distribution of parameters $\theta_i$ can be approximated by a Gaussian distribution around the true value, with covariance matrix given by the inverse of Fisher matrix: $\Sigma_{ij}=(\Gamma^{-1})_{ij}$, and
\begin{equation}
	\Gamma_{ij}=2\int_{f_\text{min}}^{f_\text{max}}\frac{\partial_ih(f)\partial_jh^\ast(f)+\partial_ih^\ast(f)\partial_jh(f)}{S_n(f)}df
\end{equation}
where $\partial_i$ is the partial derivative with respect to the $i$-th parameter of $\hat\theta=(\mc{M}_c,\eta,\chi_1,\chi_2,\ell)$. We do not take into account the polarization and inclination angle but average the orientation of the \acp{MBHB}, so the sky-averaged noise \ac{PSD} for TianQin \textcolor{black}{\cite{Wang:2019ryf}} is uesed,
\begin{eqnarray}
	S_n(f)&=&\frac{10}{3}\frac{1}{L^2}\left[1+\left(\frac{2fL_0}{0.41c}\right)^2\right]\nn\\
	&\times&\left[\frac{4S_a}{(2\pi f)^4}\left(1+\frac{10^{-4}\hz}{ f}\right)+S_x\right]
\end{eqnarray}
where $L=\sqrt{3}\scf{5}\text{km}$ is the arm length of TianQin, $\sqrt{S_a}=10^{-15}\m\dot\s^{-2}\hz^{-1/2}$ and $\sqrt{S_x}=10^{-12}\m\dot\hz^{-1/2}$ are the acceleration and position noise of TianQin, respectively.

We considered the ``3 month on + 3 month off'' observation scheme with assumption that all the events merged during the observation period. Since \ac{ppE} correction is used for the inspiral stage, we will only use the inspiral signal, and integrate frequency between
\begin{equation}
	f_\text{max}=f_\text{ISCO}=\frac{1}{6\sqrt{6}\pi M}
\end{equation}
and
\begin{equation}
	f_\text{min}=\left(\frac{5}{256}\right)^{3/8}\frac{1}{\pi}\mc{M}_c^{-5/8}T^{-3/8}.
\end{equation}

\subsection{The $F$ statistic}
Here we will introduce the statistic $F$ in \cite{yuan2024} that enables us to distinguish different effects with multiple events: by checking whether all the sources have the same parameter ${\theta_{\text{MG}}}$.

Suppose we have detected $n$ events in observations, finding deviations of \ac{GR} at some PN order, for example, here $-4$ PN order. If the modification is from modified gravity(MG) theory (in this section, the extra dimension theory \ac{RS}-II, ${\theta_{\text{MG}}}=\ell$), we would obtain the same ${\theta_{\text{MG}}}$ for all the events. However, if the modification is caused by environmental effect(here the \ac{DF} effect from \ac{DM} spike), according to \eqref{dfl}, different events will have various values depending on the source.

In either case, the true ${\theta_{\text{MG}}}$ of the $i$-th event of the $n$ detected events is ${\theta_{\text{MG}}}_i$,
with \ac{PE} precision $\delta{\theta_{\text{MG}}}_i$.
Then the posterior of ${\theta_{\text{MG}}}$ for the $i$-th event can be approximated as
a normal distribution: ${\theta_{\text{MG}}}\sim N\left({\theta_{\text{MG}}}_i,\delta{\theta_{\text{MG}}}_i\right)$.
Due to the existence of instrument noise,
the center of this distribution ${\theta_{\text{MG}}}_i^\prime$ will deviate from the true value,
which also obey this distribution ${\theta_{\text{MG}}}_i^\prime\sim N\left({\theta_{\text{MG}}}_i,\delta{\theta_{\text{MG}}}_i\right)$. The mean value of this central value of each event is
\begin{equation}
	\overline{{\theta_{\text{MG}}}^\prime}=\frac{1}{n}\sum_{i=1}^n{\theta_{\text{MG}}}_i^\prime.
\end{equation}
With the center values ${\theta_{\text{MG}}}_i^\prime$, the means of them $\overline{{\theta_{\text{MG}}}^\prime}$ and $\delta{\theta_{\text{MG}}}_i$, the statistic $F$ is obtained\cite{yuan2024}\footnote{We correct a typo in Eq. (31) of \cite{yuan2024}: the denominator there should be ${\sum_{i=1}^n\delta {\theta_{\text{MG}}}_i^2}={\sum_{i=1}^n\delta {\dot{G}}_i^2}$.},
\begin{equation}
	F=\frac{\sum_{i=1}^n\left({\theta_{\text{MG}}}^\prime_i-\overline{{\theta_{\text{MG}}}^\prime}\right)^2}{\sum_{i=1}^n\delta {\theta_{\text{MG}}}_i^2}\label{equn:stat}
\end{equation}
which characterizes the dispersion of the posteriors of all the events. When all the events have the same central values, $F=0$. However, $F$ is a small number but not exactly zero due to the random error in the \ac{PE}. On the other hand, if the central values of the $n$ events are totally different,
and the \ac{PE} precision is much smaller than the differences,
we will get a very large $F$.

\begin{figure}
	\begin{center}
		\includegraphics[scale=0.6]{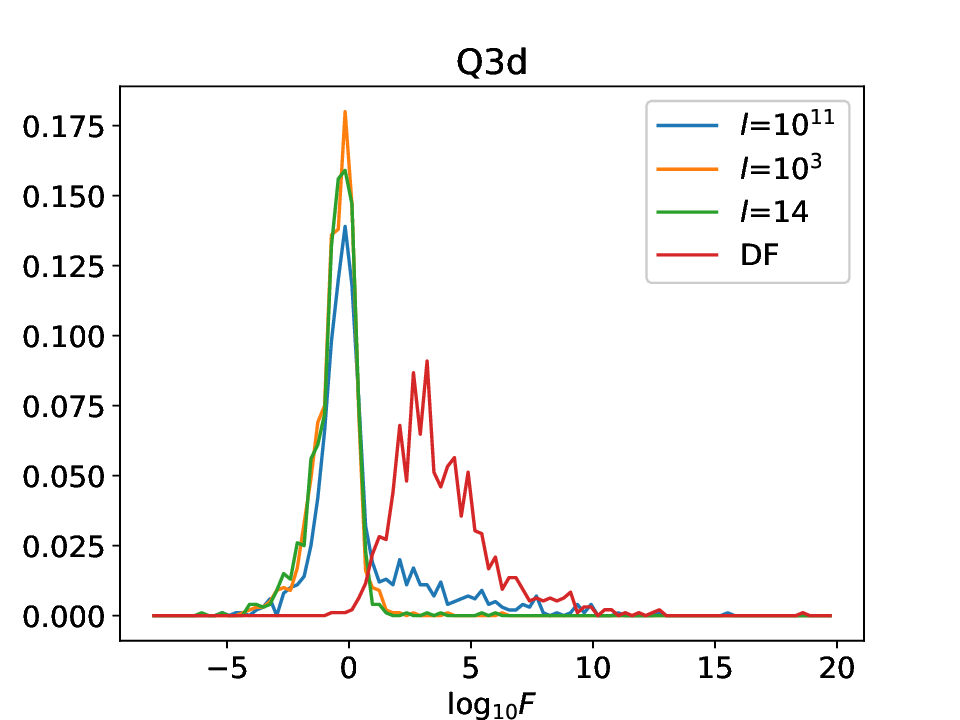}
		\caption{Comparison of the probability distributions of the $F$ statistic \eqref{equn:stat} for the Q3d source, corresponding to the \ac{DF} and fixed $\ell$ values. Here, three $\ell$ values are selected: $14\,\mu\text{m}$, $10^{3}\,\mu\text{m}$, and $10^{11}\,\mu\text{m}$.}\label{fjhq3ded}
	\end{center}
\end{figure}

\begin{figure}
	\begin{center}
		\includegraphics[scale=0.6]{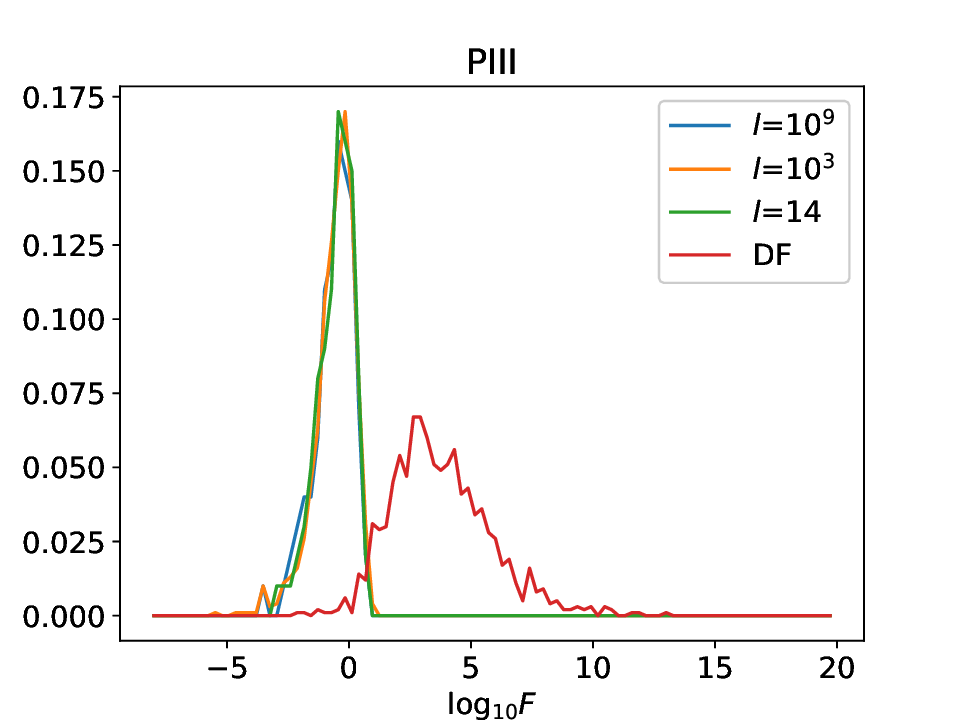}
		\caption{Comparison of the probability distributions of the $F$ statistic \eqref{equn:stat} for the PIII source, corresponding to the \ac{DF} and fixed $\ell$ values. Here, three $\ell$ values are selected: $14\,\mu\text{m}$, $10^{3}\,\mu\text{m}$, and $10^{9}\,\mu\text{m}$.}\label{fjhp3ed}
	\end{center}
\end{figure}

\begin{figure}
	\begin{center}
		\includegraphics[scale=0.6]{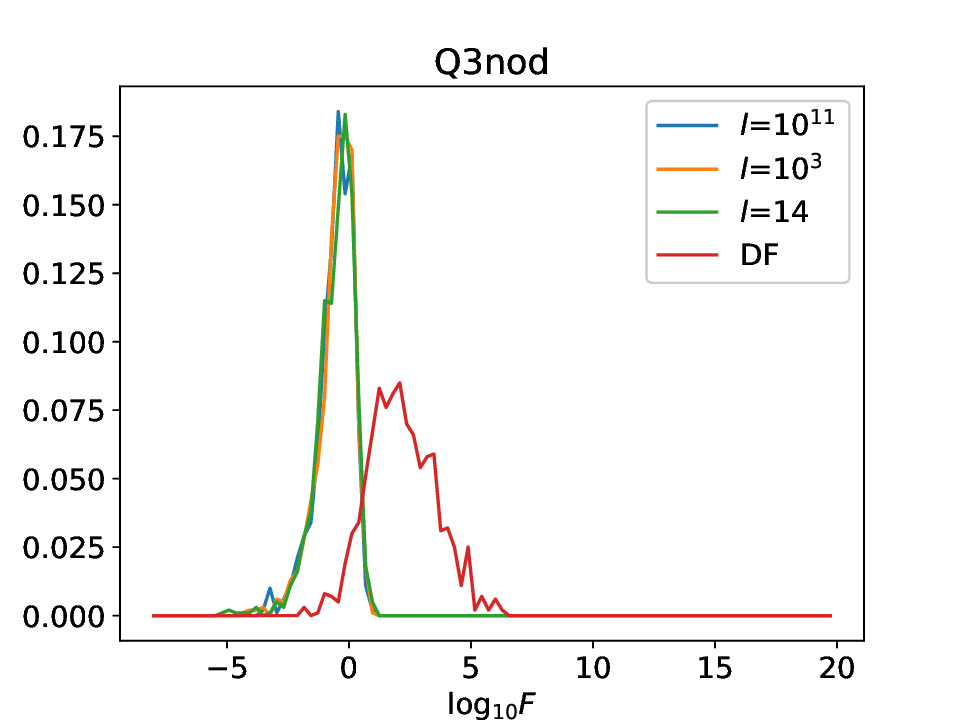}
		\caption{Comparison of the probability distributions of the $F$ statistic \eqref{equn:stat} for the Q3nod source, corresponding to the \ac{DF} and fixed $\ell$ values. Here, three $\ell$ values are selected: $14\,\mu\text{m}$, $10^{3}\,\mu\text{m}$, and $10^{11}\,\mu\text{m}$.}\label{fjhq3noded}
	\end{center}
\end{figure}

If the modifications in waveform are from \ac{DF} effect from \ac{DM} spike, the $F$ statistic \eqref{equn:stat} is calculated with central values ${\theta_{\text{MG}}}=\ell$ in \eqref{dfl} for each 1000 groups of events considering the three astronomical models introduced in Sec.~\ref{rwc}, and the distributions $F$ are ploted in \figref{fjhq3ded}, \figref{fjhp3ed} and \figref{fjhq3noded}. For comparison, we also plot the results with the same center values of $\ell$, as predicted by usual extra dimension theory. We choose three constant values of $\ell$, two are constraints mentioned before, and one is the approximate corresponding value of each model(Q3d and Q3nod: $10^{11}\mu\m$, PIII: $10^{9}\mu\m$) which can be seen from \figref{trhgded}. From \figref{fjhq3ded}, \figref{fjhp3ed} and \figref{fjhq3noded}, we observe that for all three models, \ac{DF} effect from \ac{DM} spike and extra dimension are highly distinguishable: the $F$ of \ac{DF} are generally much larger than those of extra dimension effect. The environmental effect, namely \ac{DF} effect from \ac{DM} spike, exhibits certain diffusivity due to its dependence on the dark matter density near the wave source. Correspondingly, the distribution of the extra dimension parameter $\ell$ is more diffuse, leading to a larger $F$ statistic with a wider distribution range.

This is similar to the result given by Yuan $et$ $al$\cite{yuan2024} where they use statistic $F$ to distinguish \ac{DF} effect from \ac{DM} spike and varying $G$ effect. Nevertheless, the distinguishability above is less than that in \cite{yuan2024}, and from \figref{fjhq3ded}, \figref{fjhp3ed} and \figref{fjhq3noded}, we see the distance between the two peaks of the \ac{DF} and extra dimension effects here is not as wide as that given there, the shape differences are less significant as well. Consequently, the $F$ distributions of \ac{DF} and extra dimension have more overlap than results of Yuan $et$ $al$\cite{yuan2024}, and it is not direct to find the threshold of $F$ as there. This motivates us to adopt some critieria for the threshold.

\subsection{The \ac{ROC} method}
To find a proper threshold of $F$ when distinguishing between two distributions in \figref{fjhq3ded}, \figref{fjhp3ed} and \figref{fjhq3noded}, we will employ the \ac{ROC} curve method.

The \ac{ROC} curve method can be used to examine the impact of different thresholds on distinguishing between two effects, and to identify the optimal discrimination threshold according to different purposes. The \ac{ROC} method first classifies two groups of values into positive and negative classes. By setting a threshold for the values to be distinguished, the proportion of correctly classified positive values (true positive rate, TPR), the proportion of incorrectly classified positive values (false positive rate, FPR), the proportion of correctly classified negative values (true negative rate, TNR), and the proportion of incorrectly classified negative values (false negative rate, FNR) are calculated based on whether the values are greater than or less than the threshold. By adjusting the threshold size according to the probability distributions of the two compared groups, the corresponding changes in TPR and FPR values can be obtained, thereby generating a curve of TPR versus FPR—the \ac{ROC} curve.  

\begin{figure}
	\begin{center}
		\includegraphics[scale=0.6]{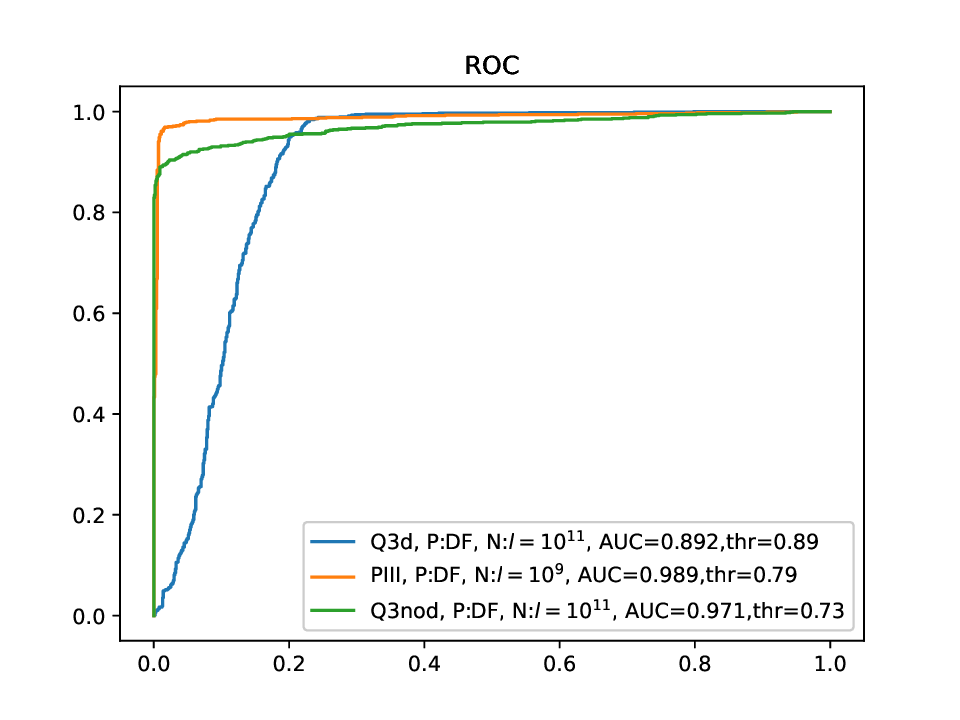}
		\caption{ROC curve for ED and fixed $\ell$ for three models.}\label{roced0}
	\end{center}
\end{figure}

For the Q3d model, the calculated ROC curve of the statistic $\log_{10}F$ for the \ac{DF} and the fixed extra dimension parameter $\ell=10^{11}\mu\m$ is shown in \figref{roced0}. To determine the optimal threshold, the Youden index method is employed, which maximizes the Youden index (TPR+TNR-1, summed rate of correctly classifying the positive and negavite classes minus 1) corresponding to the point closest to the upper-left corner of the ROC curve. The Youden index method balances sensitivity and specificity, avoiding both excessively lowering standards to pursue all positive classes and excessively raising standards to miss positive classes, which would render the threshold indistinguishable. In this way, the optimal threshold obtained is $\log F = 0.89$, in accordance with \figref{fjhq3ded}.  Thus assuming extra dimenion origin, if the $\log F$ obtained from observations with $-4$ PN deviation from \ac{GR} is smaller than $0.89$, we can say the waveform correction is really caused by extra dimension effect. However, if $\log F < 0.89$ the waveform correction is actually a result of \ac{DF} effect from \ac{DM} spike.

For the PIII and Q3nod models, the ROC curves of $\log_{10}F$ for DF and the fixed $\ell=10^{9}\mu\m$ are shown also in \figref{roced0}, with the optimal thresholds $\log F = 0.79$ and 0.73 obtained via the Youden index method, repectively.  

\begin{figure}
	\begin{center}
		\includegraphics[scale=0.6]{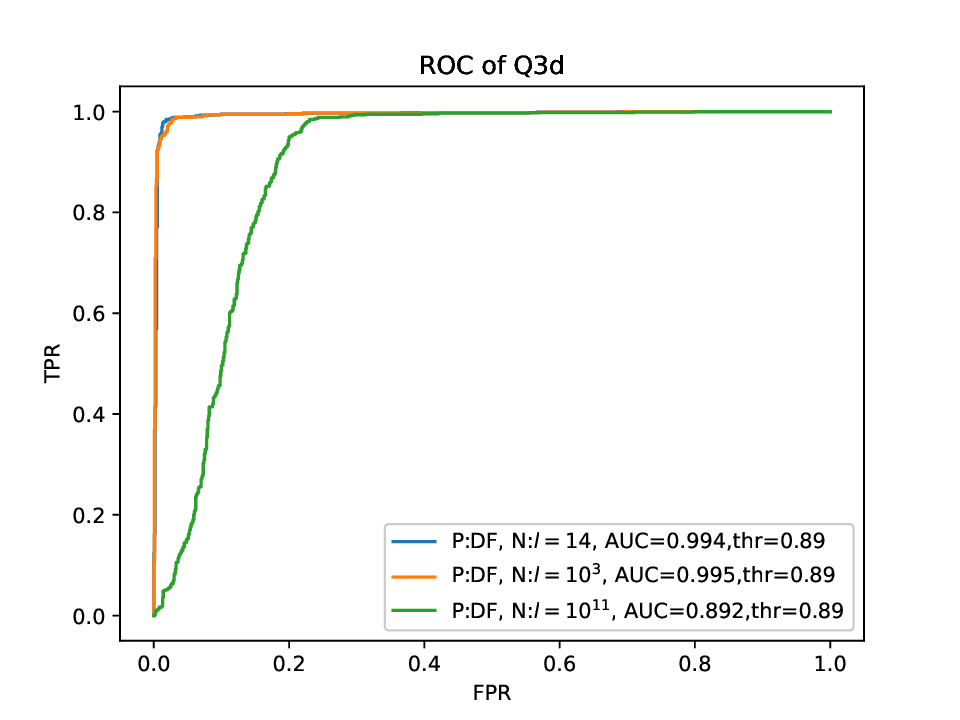}
		\caption{ROC curves between DF and different fixed $\ell$ extra dimension effects in the Q3d source.}\label{rocedcz}
	\end{center}
\end{figure}

For a specific model, the fixed $\ell$ value can be altered instead of using the value roughly corresponding to DF. For the Q3d model, the ROC curve results for the $\log F$ statistics of \ac{DF} effect from \ac{DM} spike and the extra dimension effect with another fixed $\ell$ are shown in \figref{rocedcz}. As previously mentioned, using the Youden index method, the optimal threshold between DF and the fixed $\ell=10^{11}\mu\m$ extra dimension effect is 0.89, with an \ac{AUC} of 0.892. For the other two constant values $\ell=14\mu\m$ and $10^3\mu\m$, the optimal threshold remains 0.89, and the \ac{AUC} values are all very close to 1. For the Q3d model, when the constant $\ell$ is not set to the value roughly corresponding to DF, the \ac{AUC} value is higher (indicating better discrimination between the two effects), though this has no impact on threshold selection.  

\begin{figure}
	\begin{center}
		\includegraphics[scale=0.6]{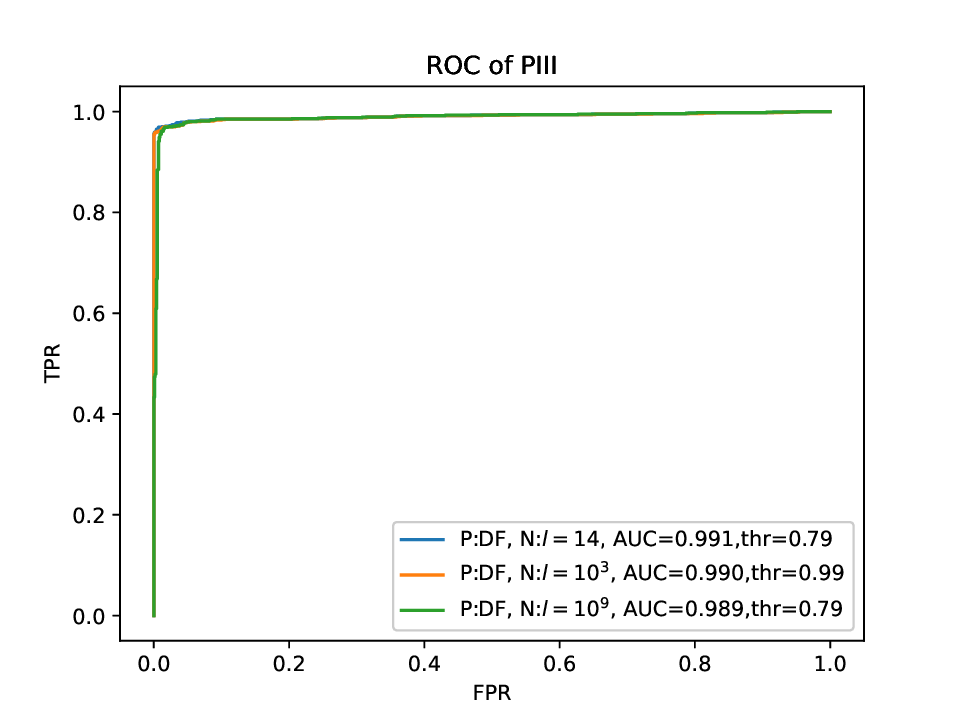}
		\caption{ROC curves between DF and different fixed $\ell$ extra dimension effects in the PIII source.}\label{rocedczp3}
	\end{center}
\end{figure}

For the PIII model, the ROC curve results for the $\log F$ statistics of the two effects with different fixed $\ell$ are shown in \figref{rocedczp3}. The optimal threshold between DF and the fixed $\ell=10^9\mu\m$ extra dimension effect is 0.89, with an \ac{AUC} of 0.989. For $\ell=14\mu\m$ and $10^3\mu\m$, the optimal thresholds are 0.79 and 0.99, respectively, with \ac{AUC} values close to 1 (0.991 and 0.990). For the PIII model, when the constant $\ell$ is not set to the value corresponding to DF, the \ac{AUC} value is slightly higher (better discrimination), with minimal impact, though the threshold selection changes slightly.  

\begin{figure}
	\begin{center}
		\includegraphics[scale=0.6]{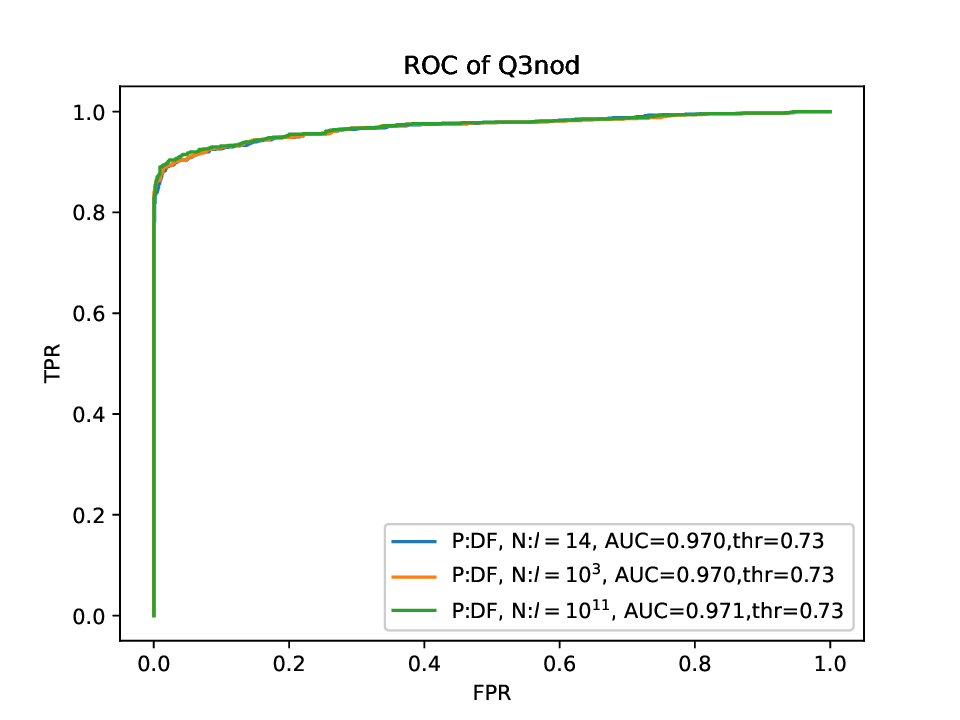}
		\caption{ROC curves between DF and different fixed $\ell$ extra dimension effects in the Q3nod source.}\label{rocedczq3nod}
	\end{center}
\end{figure}

For the Q3nod model, the ROC curve results for the $\log F$ statistics of the two effects are shown in \figref{rocedczq3nod}. The optimal threshold between DF and the fixed $\ell=10^{11}\mu\m$ extra dimension effect is 0.73, with an \ac{AUC} of 0.971. For $\ell=14\mu\m$ and $10^3\mu\m$, the optimal threshold remains 0.73, and the \ac{AUC} values are both 0.970. For the Q3nod model, the discrimination degree (\ac{AUC} value) between the two effects is insensitive to $\ell$, and the optimal threshold remains unchanged.  

For briefness, we summarize these results in \tabref{tab:1}. If we obtain $\log_{10}F>\tx{thr}$ from observations, then \ac{DF} effect from \ac{DM} spike is preferred to be true, otherwise extra dimension effect is.
\begin{table*}
	\caption{The thresholds of $\log_{10}F$ and \ac{AUC}: $(\tx{thr},\tx{\ac{AUC}})$, setting \ac{DF} is positive for different models, obtained from  \figref{rocedcz}, \figref{rocedczp3} and \figref{rocedczq3nod}.}
	\label{tab:1}       
	\centering
	\begin{tabular}{c|ccc}
		\hline\noalign{\smallskip}
		$(\tx{thr},\tx{\ac{AUC}})$ & $\ell=14$ & $\ell=10^3$ & $\ell=\ell_\tx{c}$\\
		\noalign{\smallskip}\hline\noalign{\smallskip}
		Q3d &$(0.89, 0.994)$ & $(0.89, 0.995)$ &$\ell_\tx{c}=10^{11}$: $(0.89, 0.892)$\\
		PIII &$(0.79, 0.991)$ & $(0.99, 0.990)$ &$\ell_\tx{c}=10^{9}$: $(0.79, 0.989)$\\
		Q3nod &$(0.73, 0.970)$ & $(0.73, 0.970)$ &$\ell_\tx{c}=10^{11}$: $(0.73, 0.971)$\\
		\noalign{\smallskip}\hline
	\end{tabular}
	\vspace*{5cm}  
\end{table*}

\section{Distinguishing between varying $G$ theory, extra dimension, and \ac{DF} from \ac{DM} spike effects}\label{sec:4}

The procedures proposed by Yuan $et$ $al$\cite{yuan2024} and added above can also be applied to the distinguishment between other effects at the same PN order, as long as ratios of their concerned parameters are not completely a constant independent of source. In this section we will use the $F$ statistic for the distinguishment between the three effects all at $-4$ PN order: extra dimension effect, varying $G$ effect and \ac{DF} effect from \ac{DM} spike. The former two effects are distinguished in \secref{sec:4.1}, and later together with the result in \cite{yuan2024}, these three effects will be compared directly with the statistic $F$, which means the range of $F$ in future observations can pick out the right effect immediately(for a specific astronomical model).

\subsection{Distinguishing extra dimension and  varying $G$ effects}\label{sec:4.1}
The analysis in \cite{yuan2024} have implemented the distinguishment between \ac{DF} induced $\dot{G}(\rho_0)$ distribution (from \eqref{dfxz} and \eqref{sbyl}) of sources and usual constant value $\dot{G}$ using the statistic $F$, choosing ${\theta_{\text{MG}}}=\dot{G}$ in \eqref{equn:stat}. Similar distinguishment can be done between extra dimension induced $\dot{G}(\rho_0)$ distribution of sources and usual constant value $\dot{G}$. 

\begin{figure}
	\begin{center}
		\includegraphics[scale=0.55]{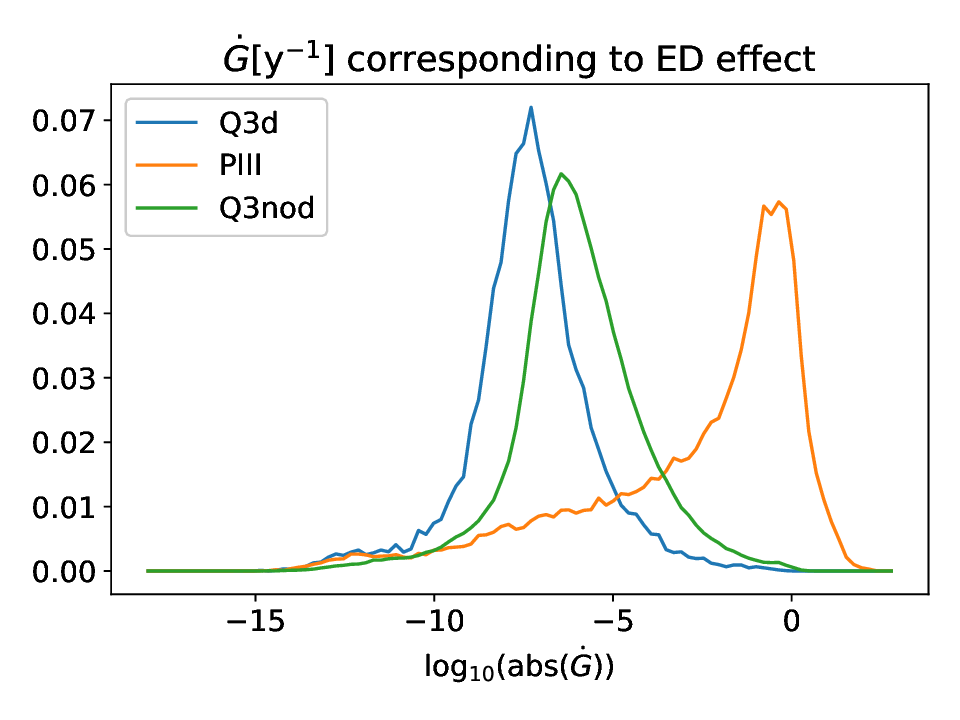}
		\caption{$\dot{G}$ distributions for extra dimension effects of Q3d, PIII, and Q3nod sources derived from Equation \eqref{edgdzh}.}\label{tedgd}
	\end{center}
\end{figure}

Since the waveform correction of the extra dimension effect also shares the \ac{ppE} parameter $ b = -13 $ with the varying $G$ effect, equating their phase expressions \eqref{edxw} and \eqref{sbyl} yields the $\dot{G}$ value corresponding to the extra dimension theory parameter $\ell$ as:  
\ba
\dot{G}(\ell) &=& -1.3 \times 10^{-24} \left( \frac{M_\odot}{m_1^2} + \frac{M_\odot}{m_2^2} \right) \left( \frac{\ell}{10\mu\text{m}} \right)^2\nn\\
&\times&\frac{3 - 26 \eta + 34 \eta^2}{\eta^{\frac{2}{5}} (1 - 2 \eta)} \frac{131072}{25 \mc{M}_c} \label{edgdzh}
\ea
Based on this, the $\dot{G}$ distributions for the extra dimension effects of the three astronomical models (Q3d, PIII, and Q3nod sources) are shown in \figref{tedgd}, where $\ell = 10^9 \mu\text{m}$ is selected. Since Equation \eqref{edgdzh} has a zero point at $\eta \approx 0.14$, $\dot{G}$ is positive for $\eta \ge 0.14$ and negative for most other values. In \figref{tedgd}, the absolute value of $\dot{G}$ is taken. As shown, the PIII source exhibits a significantly different distribution from the other two models, with its peak near $\dot{G} = -10^{-1}\text{y}^{-1}$, because Equation \eqref{edgdzh} decreases rapidly with increasing mass, and PIII contains a higher proportion of low-mass sources. The peaks for Q3d and Q3nod sources are at $\dot{G} = -10^{-8}\text{y}^{-1}$ and $\dot{G} = -10^{-7}\text{y}^{-1}$, respectively.  

\begin{figure}
	\begin{center}
		\includegraphics[scale=0.6]{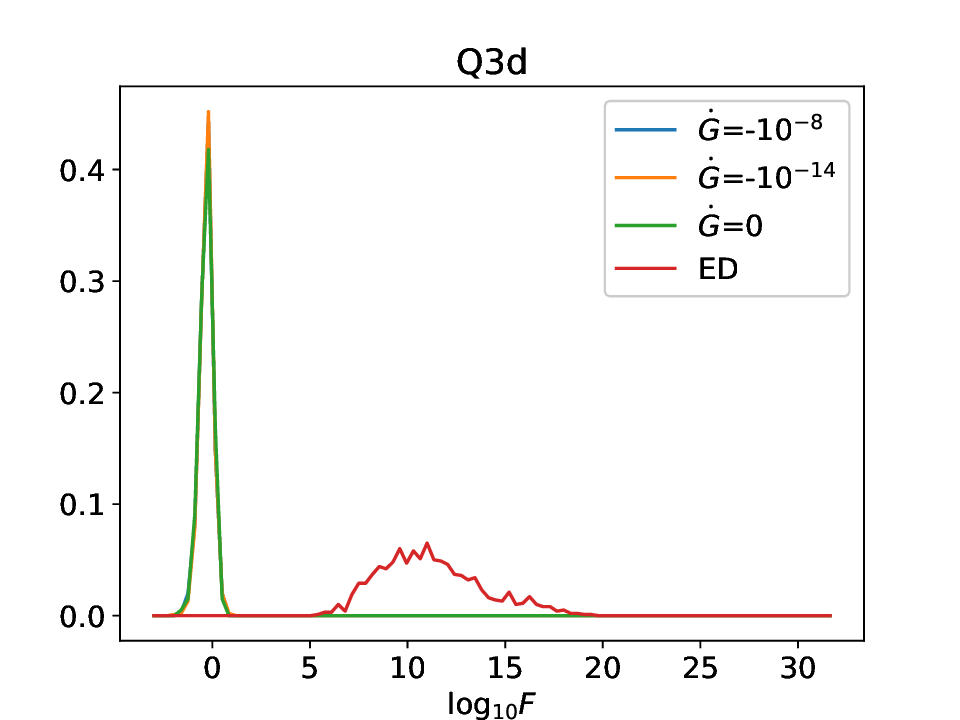}
		\caption{Comparison of probability distributions of the $F$ statistic \eqref{equn:stat} for Q3d source, corresponding to ED and fixed $\dot{G}$ values. The fixed $\dot{G}$ values selected here are $0$, $-10^{-8}\text{y}^{-1}$, and $-10^{-14}\text{y}^{-1}$.}\label{fjhq3dedgd}
	\end{center}
\end{figure}

\begin{figure}
	\begin{center}
		\includegraphics[scale=0.6]{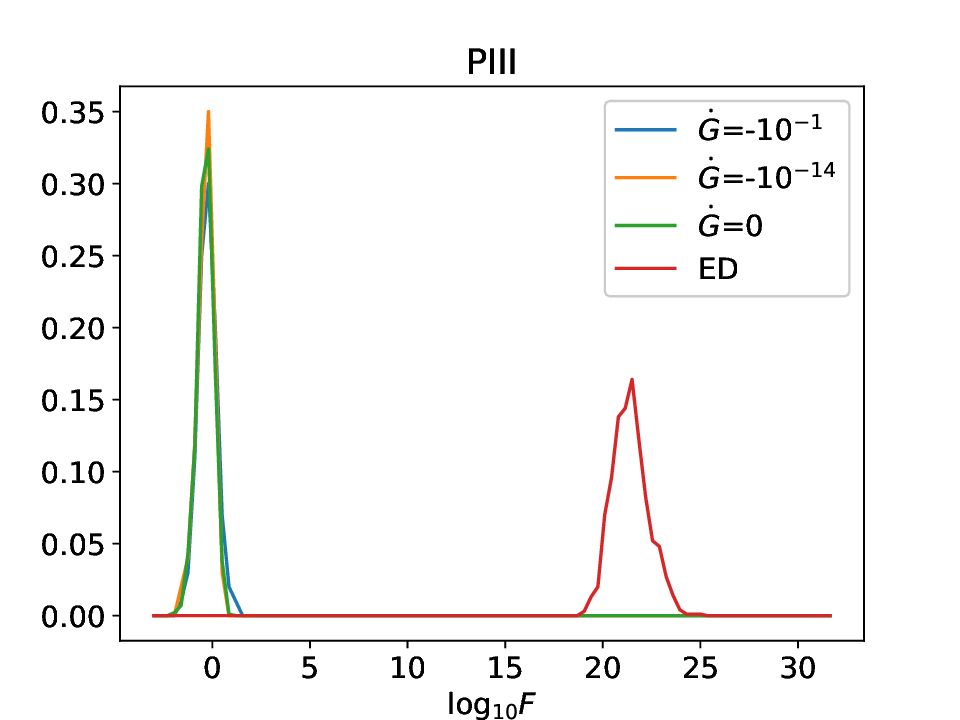}
		\caption{Comparison of probability distributions of the $F$ statistic \eqref{equn:stat} for PIII source, corresponding to ED and fixed $\dot{G}$ values. The fixed $\dot{G}$ values selected here are $0$, $-10^{-1}\text{y}^{-1}$, and $-10^{-14}\text{y}^{-1}$.}\label{fjhp3edgd}
	\end{center}
\end{figure}

\begin{figure}
	\begin{center}
		\includegraphics[scale=0.6]{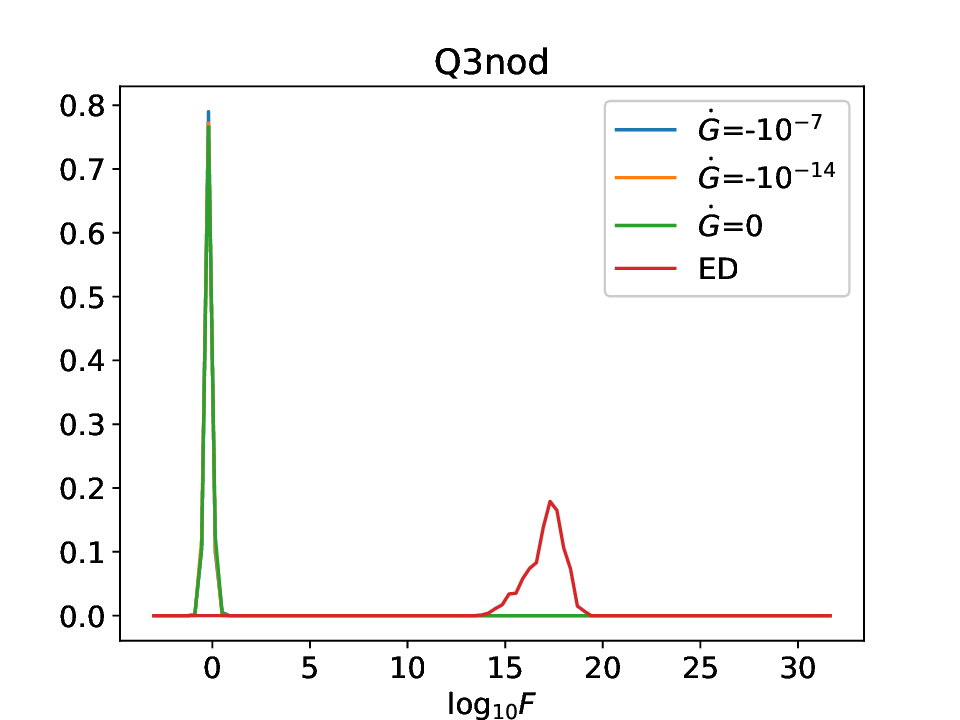}
		\caption{Comparison of probability distributions of the $F$ statistic \eqref{equn:stat} for Q3nod source, corresponding to ED and fixed $\dot{G}$ values. The fixed $\dot{G}$ values selected here are $0$, $-10^{-7}\text{y}^{-1}$, and $-10^{-14}\text{y}^{-1}$.}\label{fjhq3nodedgd}
	\end{center}
\end{figure}

Next, multi-event simulations are performed for the $\dot{G}$ values corresponding to the three astronomical sources and the conventional varying $G$ effect, with the calculated $F$ statistic results shown in \figref{fjhq3dedgd}, \figref{fjhp3edgd}, and \figref{fjhq3nodedgd}. In addition to the strongest current constraint on $\dot{G}$, the respective $\dot{G}$ peaks from \figref{tedgd} are also selected. Among the three models, the PIII source yields the largest $F$ statistic, followed by Q3nod and Q3d. This is because the PIII source’s higher number of sources with asymmetric mass ratio enhances the varying $G$ effect, leading to a smaller $\dt\dot{G}$ and thus a larger $F$ value. Additionally, the more diffuse mass distribution results in a wider $F$ value range.  

\begin{figure}
	\begin{center}
		\includegraphics[scale=0.6]{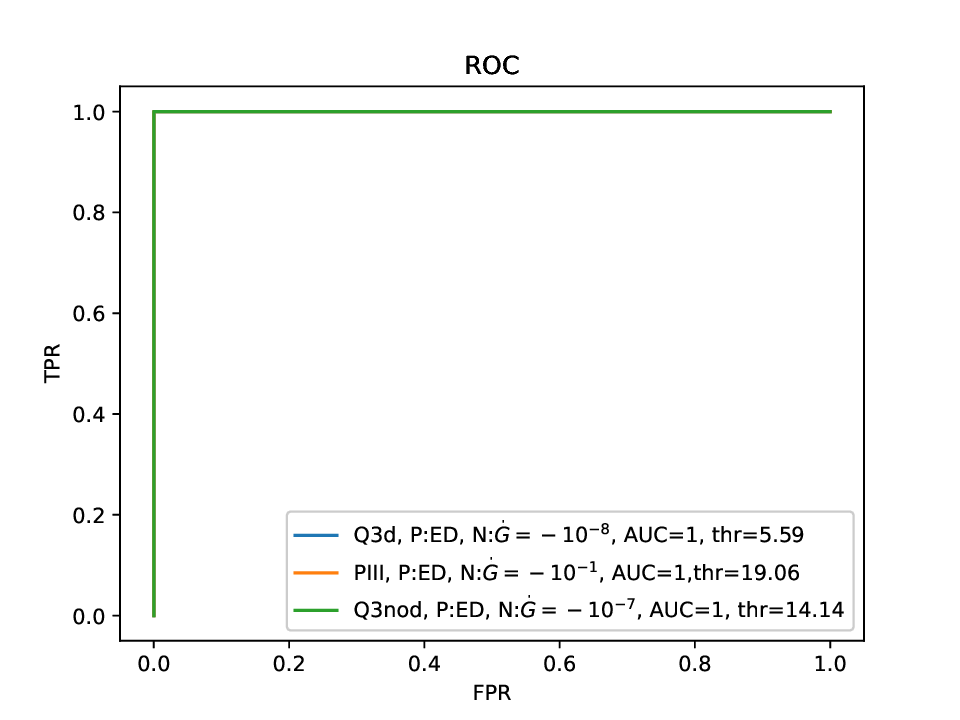}
		\caption{ROC curve for ED and fixed $\dot{G}$ for the three models.}\label{rocedgd7}
	\end{center}
\end{figure}

Again, the ROC curve method can further distinguish between these effects. For the Q3d model, the result is shown in \figref{rocedgd7}, with the optimal threshold $\log F = 5.59$ obtained via the Youden index method.  
For the PIII model, the optimal threshold $\log F = 19.06$.  
For the Q3nod model, the optimal threshold $\log F = 14.14$.  

\subsection{Distinguishing between the three effects}  
\begin{figure}
	\begin{center}
		\includegraphics[scale=0.6]{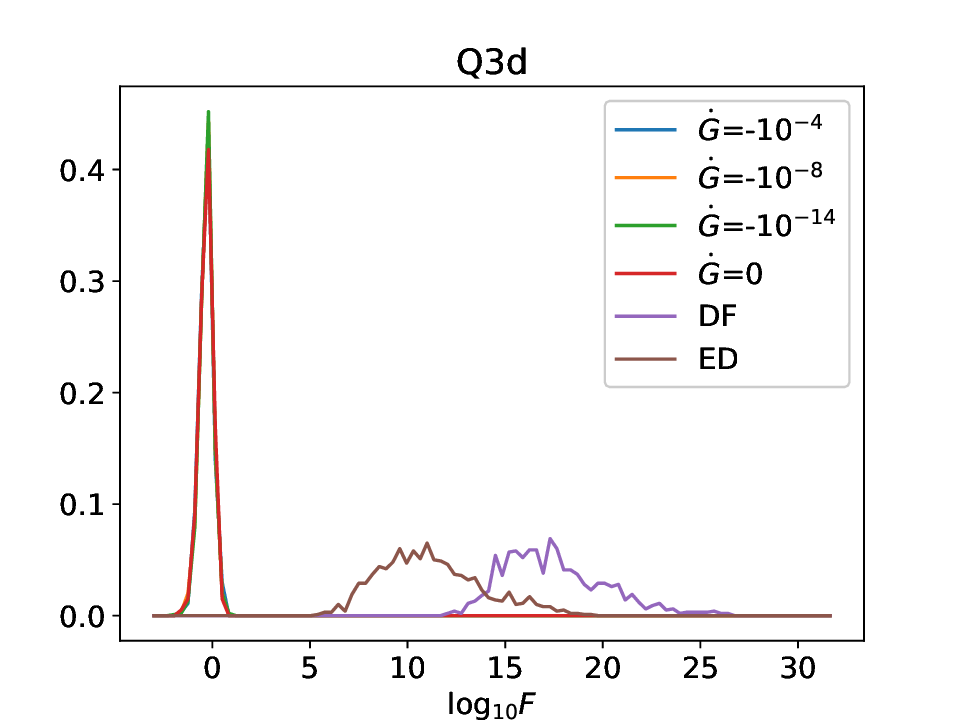}
		\caption{Comparison of probability distributions of the $F$ statistic \eqref{equn:stat} for Q3d source, corresponding to DF, ED, and fixed $\dot{G}$ values.}\label{fjhq3dedgddf}
	\end{center}
\end{figure}

\begin{figure}
	\begin{center}
		\includegraphics[scale=0.6]{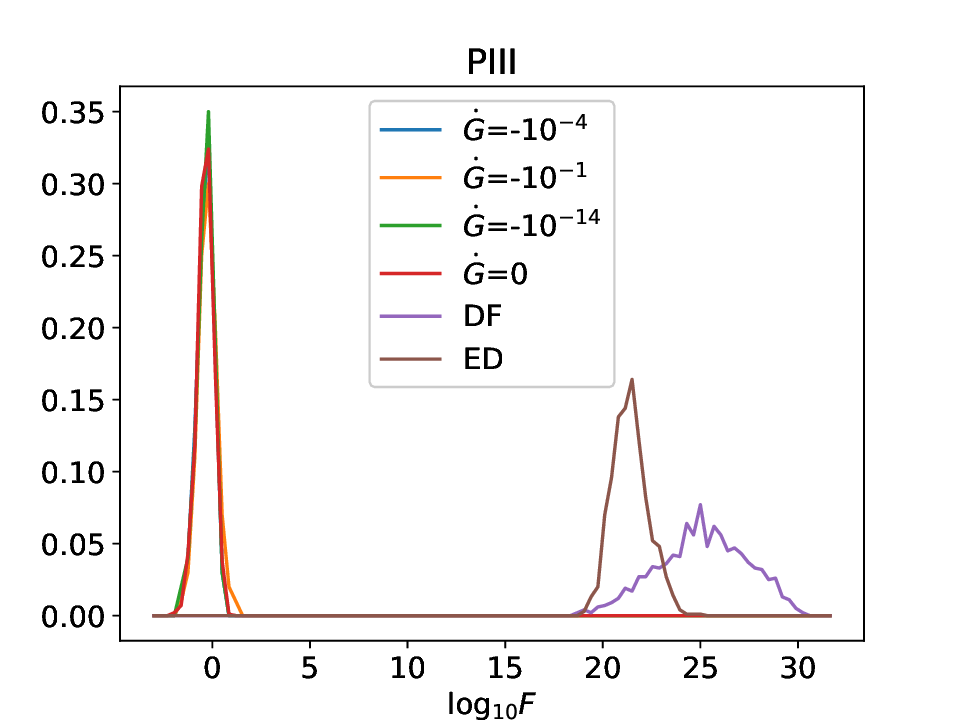}
		\caption{Comparison of probability distributions of the $F$ statistic \eqref{equn:stat} for PIII source, corresponding to DF, ED, and fixed $\dot{G}$ values.}\label{fjhp3edgddf}
	\end{center}
\end{figure}

\begin{figure}
	\begin{center}
		\includegraphics[scale=0.6]{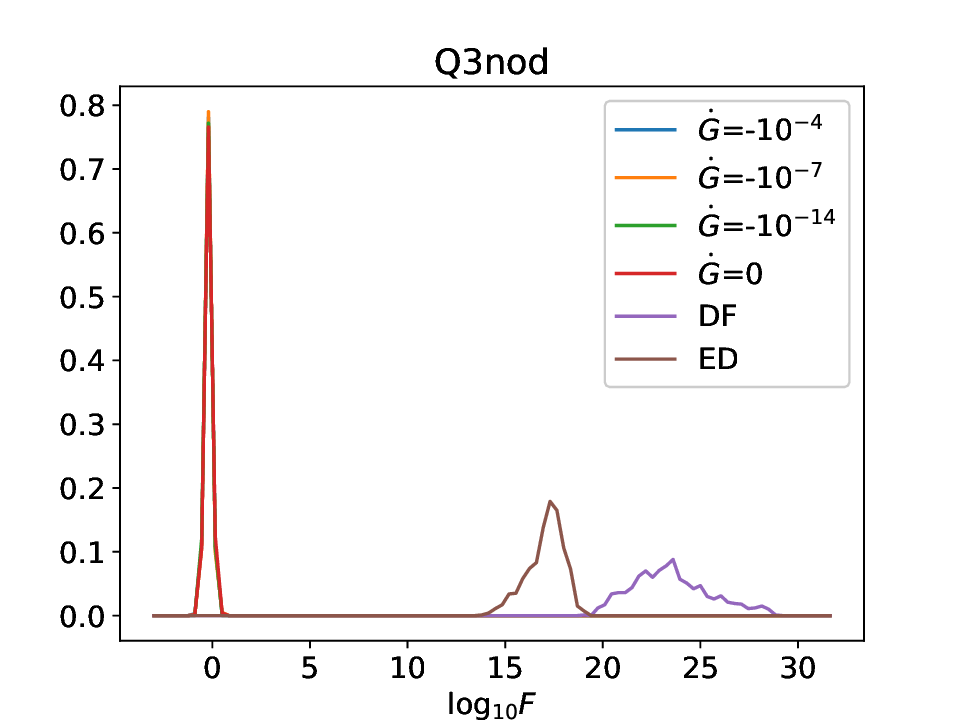}
		\caption{Comparison of probability distributions of the $F$ statistic \eqref{equn:stat} for Q3nod source, corresponding to DF, ED, and fixed $\dot{G}$ values.}\label{fjhq3nodedgddf}
	\end{center}
\end{figure}

Combining the $F$ comparison between $\dot{G}$ and \ac{DF} effect in FIG.8-10 of \cite{yuan2024} with comparison between $\dot{G}$ and extra dimension in \figref{fjhq3dedgd}, \figref{fjhp3edgd} and \figref{fjhq3nodedgd}, we get \figref{fjhq3dedgddf}, \figref{fjhp3edgddf}, and \figref{fjhq3nodedgddf}. For convenience, both the \ac{DF} effect from \ac{DM} spike and extra dimension are conversed to correponding $\dot{G}$. For fixed $\dot{G}$ values, we include not only the existing values $0$, $-10^{-4}\text{y}^{-1}$, and $-10^{-14}\text{y}^{-1}$ (corresponding to GR, DF, and the strongest constraint) but also the respective $\dot{G}$ peaks of the extra dimension effect for the three astronomical models.  

\begin{figure}
	\begin{center}
		\includegraphics[scale=0.6]{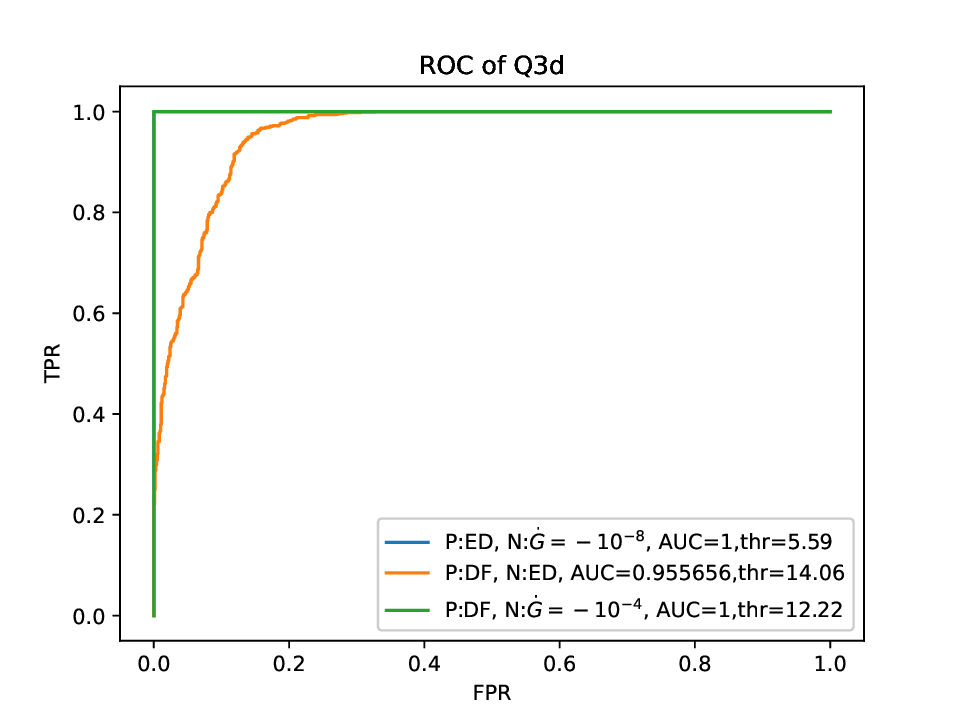}
		\caption{ROC curves between DF, ED, and fixed $\dot{G}$ effects in the Q3d source.}\label{rocedgd7df}
	\end{center}
\end{figure}

For the Q3d model, the ROC curve results comparing the $\log F$ values of the \ac{DF} effect from \ac{DM} spike, varying $G$ effect, and extra dimension effect are shown in \figref{rocedgd7df}. As previously mentioned, the optimal threshold between extra dimension and fixed varying $G$ effects is 5.59 via the Youden index method. Results for the other two effect pairs are as follows: the optimal threshold between extra dimension and \ac{DF} effects is 14.06, and between \ac{DF} and fixed varying $G$ effects is 12.22.  

\begin{figure}
	\begin{center}
		\includegraphics[scale=0.6]{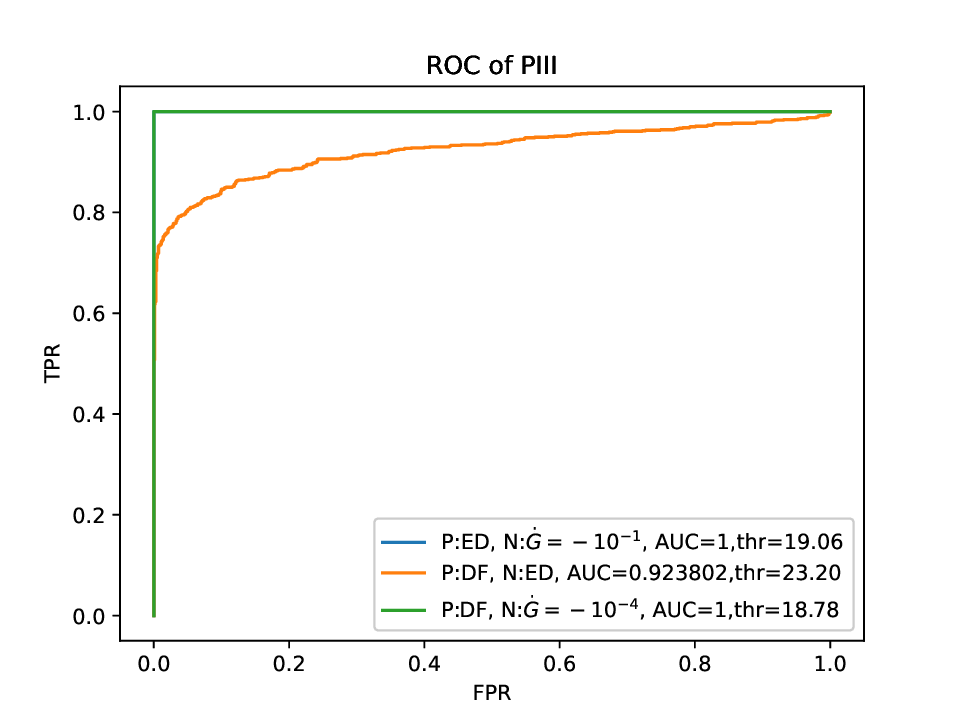}
		\caption{ROC curves between DF, ED, and fixed $\dot{G}$ effects in the PIII source.}\label{rocedgd7dfp3}
	\end{center}
\end{figure}

For the PIII model, the ROC curve results for the three effects are shown in \figref{rocedgd7dfp3}. The optimal threshold between extra dimension and fixed varying $G$ effects is 19.06. The optimal thresholds between extra dimension and \ac{DF} effects is 23.20, and between \ac{DF} and fixed varying $G$ effects is 18.78.  

\begin{figure}
	\begin{center}
		\includegraphics[scale=0.6]{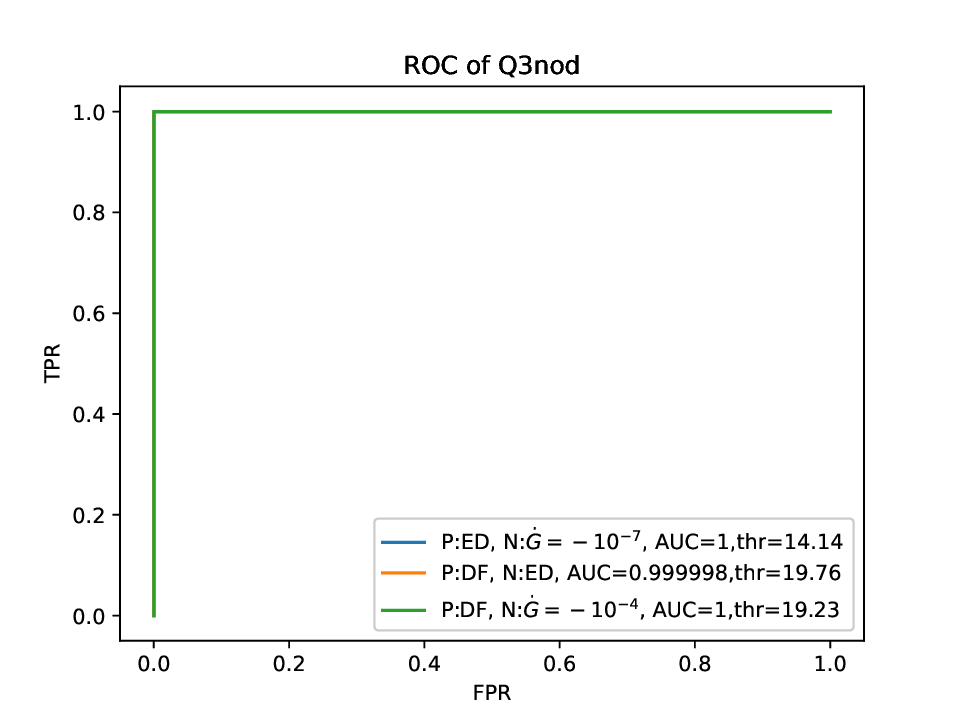}
		\caption{ROC curves between DF, ED, and fixed $\dot{G}$ effects in the Q3nod source.}\label{rocedgd7dfq3nod}
	\end{center}
\end{figure}

For the Q3nod model, the ROC curve results are shown in \figref{rocedgd7dfq3nod}. The optimal threshold between extra dimension and fixed varying $G$ effects is 14.14. The optimal thresholds between extra dimension and \ac{DF} effects is 19.76, and between \ac{DF} and fixed varying $G$ effects is 19.23.  

From \figref{rocedgd7df}, \figref{rocedgd7dfp3} and \figref{rocedgd7dfq3nod}, after conversing \ac{DF} and ED effects to $\dot{G}$ effect, the main results of the distingushing thresholds of statistic $F$ between three effects are collected in \tabref{tab:2} with the last columm specifying former results obtained by Yuan $et$ $al$\cite{yuan2024}.
\begin{table*}
	\caption{The thresholds of $\log_{10}F$ and \ac{AUC}: $(\tx{thr},\tx{\ac{AUC}})$ between three effects for different models, obtained from  \figref{rocedgd7df}, \figref{rocedgd7dfp3} and \figref{rocedgd7dfq3nod}.}
	\label{tab:2}       
	\centering
	\begin{tabular}{c|ccc}
		\hline\noalign{\smallskip}
		$(\tx{thr},\tx{\ac{AUC}})$ & P: ED, N: $\dot{G}$ & P: \ac{DF}, N: ED  & P: \ac{DF}, N: $\dot{G}$\\
		\noalign{\smallskip}\hline\noalign{\smallskip}
		Q3d&$\dot{G}_\tx{c}=-10^{-8}: (5.59, 1)$ & $(14.06, 0.955656)$ & $\dot{G}_\tx{c}=-10^{-4}: (12.22, 1)$\\
		PIII&$\dot{G}_\tx{c}=-10^{-1}: (19.06, 1)$ & $(23.20, 0.923802)$ & $\dot{G}_\tx{c}=-10^{-4}: (18.78, 1)$\\
		Q3nod&$\dot{G}_\tx{c}=-10^{-7}: (14.14, 1)$ & $(19.76, 0.999998)$ & $\dot{G}_\tx{c}=-10^{-4}: (19.23, 1)$\\
		\noalign{\smallskip}\hline
	\end{tabular}
	\vspace*{5cm}  
\end{table*}

To sum up, in future observation if $-4$ PN order waveform deviation of \ac{GR} are detected in \acp{GW} from \ac{MBHB}, the statistic $F$ can be calculated to pick out the right effect with a considered astronomical model. Generally speaking, \ac{DF} effect has the largest $F$ while varying G effect has the smallest, so if the detected $\log_{10}F<5.59$, it must come from varying G effect and if $\log_{10}F>23.20$, \ac{DF} effect from \ac{DM} spike. If we detected other $F$, the result will depend on which astronomical model is selected and the corresponding threshold in \tabref{tab:2}, \figref{fjhq3dedgddf}, \figref{fjhp3edgddf}, and \figref{fjhq3nodedgddf}. Other effects at $-4$ PN order, for example, Bondi-Hoyle accretion of matter to \ac{BH}\cite{Barausse:2014tra,constr} or migration torque of disk with some $\gamma$ value\cite{kocsis,Speri:2022upm,yunes} also can be included with present results. 

\section{Conclusions}\label{concl}
In this article, we follow the method proposed by Yuan $et$ $al$\cite{yuan2024} again to distinguish environmental effect(\ac{DF} effect from \ac{DM} spike with $\gamma={3\over 2}$) in \ac{GW} corrections of \ac{MBHB} inspiral from modified theory effect of gravity(extra dimension theory of \ac{RS}-II). First we derive the relation between \ac{DM} spike density parameter $\rho_0$ and the size $\ell$ of large extra dimension in theory of \ac{RS}-II) when both effects introduce \ac{GW} phase correction at $-4$ PN order. Then we carry out simulated events based on catalog of Q3d, PIII and Q3nod models, for $\ell$ distributions of usual constant case of the extra dimension theory and for the distribution conversed through the conversion relation. We adopt the statistic $F$ to describe the dispersion of measured $\ell$ for the simulated events and find $F$ is much larger if the deviation is caused by constant $\ell$ predicted by usual extra dimension theory than originating from \ac{DM} \ac{DF} effect. So two effects are highly distinguishable. The result is similar to that in former distinguishment between \ac{DM} \ac{DF} and varying $G$ effects\cite{yuan2024} but with more overlap of $F$ distribution between \ac{DF} and extra dimension effects. Since it is not direct to find a threshold like in\cite{yuan2024}, we apply the \ac{ROC} curve method to see how the true positive rate and false positive rate vary with different thresholds and determine a proper shreshold of $\log_{10}F$ arround 0.8 which maximizes the Youden index. When choosing other constant $\ell$ values for usual extra dimension effect, the determined threshold is slightly affected, but making the \ac{AUC} approaching to 1.

Then we extend the work of Yuan $et$ $al$\cite{yuan2024} to do the distinguishment between three effects all introducing $-4$ PN order correction: \ac{DF} from \ac{DM} spike, extra dimension theory and varying $G$ theory. We implement the simulated multiple event process considering constant $\dot{G}$ value of usual varying $G$ theory, $\dot{G}$ corresponding to \ac{DF} from \ac{DM} spike and corresponding to extra dimension effect and derive the statistic $F$, respectively. The distinguishing thresholds between each two effects are then given by the \ac{ROC} curve method as before. It turns out that varying $G$ effect is more different from the other two effects, while the difference between the \ac{DM} effect and extra dimension effect is minor thus corresponding to smaller \ac{AUC}. According to the thresholds obtained, in future observations we can determine the accounting effect from detected statistic $F$ directly. Furthermore, other effects introduced at $-4$ PN order correction like Bondi-Hoyle accretion and migration torque effects can also be added.

Although the central values of $\dot{G}$ and $\ell$ corresponding to \ac{DF} from \ac{DM} spike are both much larger than current best constraints, the forementioned process can be applied to the distinguishment among other environmental effects and modified theories of gravity with \ac{GW} correction at other PN order. For example, both the gravity effect from \ac{DM} spike and \ac{EA} theory have the same order $-1$ PN order \ac{GW} correction and they can be analyzed with the $F$ statistic for observed events.


\bibliography{references}

\end{document}